\documentclass[10pt]{article}

\usepackage{arxiv}
\usepackage{iftex}
\usepackage[T1]{fontenc}
\usepackage{mathptmx}
\usepackage[scaled=0.92]{helvet}
\usepackage{courier}
\usepackage{amsmath,amssymb,amsthm}
\usepackage{microtype}
\usepackage{booktabs}
\usepackage{tabularx}
\usepackage{array}
\usepackage{enumitem}
\usepackage{graphicx}
\usepackage{listings}
\usepackage{xcolor}
\usepackage{tikz}
\usetikzlibrary{arrows.meta,positioning,fit,backgrounds,calc,shapes.geometric}
\ifXeTeX\XeTeXdefaultencoding "latin1"\fi
\usepackage[ruled,vlined,linesnumbered]{algorithm2e}
\ifXeTeX\XeTeXdefaultencoding "utf8"\fi
\usepackage[numbers,sort&compress]{natbib}
\usepackage{xurl}
\usepackage[colorlinks=true,linkcolor=blue!50!black,citecolor=blue!50!black,urlcolor=blue!50!black]{hyperref}
\hypersetup{
  pdftitle={From Conversations to Commitments: A Design Study of Shared Records for Agent-to-Business Coordination},
  pdfauthor={Carlos Eberhardt and Mansura Habiba},
  pdfsubject={Shared records, commercial confidentiality and agent-to-business coordination},
  pdfkeywords={Agent-to-business commerce, AT Protocol, delegated authority, distributed commitment, information-flow control, multi-agent systems},
  pdfdisplaydoctitle=true,
  bookmarksnumbered=true
}
\usepackage[capitalise,noabbrev]{cleveref}
\usepackage{caption}
\crefname{algocf}{Algorithm}{Algorithms}
\Crefname{algocf}{Algorithm}{Algorithms}
\newtheorem{proposition}{Proposition}

\setlist{itemsep=2pt,topsep=3pt,parsep=0pt}

\newcolumntype{L}{>{\raggedright\arraybackslash}X}
\newcolumntype{P}[1]{>{\raggedright\arraybackslash}p{#1}}

\SetKwInput{KwInput}{Input}
\SetKwInput{KwState}{State}
\SetKwInput{KwOutput}{Output}
\SetKwProg{Proc}{procedure}{}{end}
\DontPrintSemicolon
\SetAlFnt{\small}

\newcommand{\nsid}[1]{\texttt{#1}}

\definecolor{inherit}{RGB}{218,232,252}
\definecolor{added}{RGB}{255,236,204}
\definecolor{edgegray}{RGB}{110,110,110}
\definecolor{kwblue}{RGB}{0,70,140}
\definecolor{strgreen}{RGB}{30,110,50}

\lstdefinelanguage{json}{
  morestring=[b]",
  stringstyle=\color{strgreen},
  literate=
    {:}{{{\color{kwblue}{:}}}}{1}
    {,}{{{\color{kwblue}{,}}}}{1}
    {\{}{{{\color{kwblue}{\{}}}}{1}
    {\}}{{{\color{kwblue}{\}}}}}{1}
    {[}{{{\color{kwblue}{[}}}}{1}
    {]}{{{\color{kwblue}{]}}}}{1},
}
\title{From Conversations to Commitments:\\ A Design Study of Shared Records for Agent-to-Business Coordination}

\renewcommand{\shorttitle}{From Conversations to Commitments}
\renewcommand{\undertitle}{A Preprint}

\author{%
  \textbf{Carlos Eberhardt}\\
  Technical Product Management\\
  IBM Software\\
  Rochester, United States of America\\
  \texttt{carlos.eberhardt@gmail.com}
  \And
  \href{https://orcid.org/0000-0001-9051-1370}{\textcolor{black}{\includegraphics[scale=0.06]{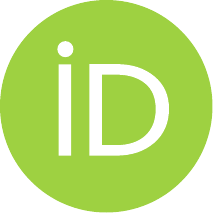}\hspace{1mm}\textbf{Mansura Habiba}}}\\
  Principal Platform Architect\\
  IBM Software\\
  Dublin, Ireland\\
  \texttt{mansurah.habiba@gmail.com}
}

\date{November 2026}

\begin{document}
\maketitle

\begin{abstract}
Business commitments require accepted terms to be linked to delegated authority, reserved resources and durable evidence. We examine how a shared, attributable interaction record can support these links across independently operated organisations. Talea combines signed AT Protocol records with A2A task execution. We use its coordination sandbox and a proposed commercial extension in a design study of an airline booking disrupted passengers into three competing hotels. Four observations follow. First, AT Protocol provides mechanisms supporting five of six coordination requirements, while record-level access and release controls require additional enforcement. Second, the case's confidentiality constraints restrict each hotel's authorised view to its bilateral exchange; only the airline can correlate all three deals. Third, this leaves a possible benefit for late-joining participants inside the airline's trust boundary, but does not establish an advantage over an equally instrumented point-to-point design. Fourth, a deterministic release interface constrains fixed, approved payloads but cannot bound disclosure when negotiated values depend on private state. Finite-model explorations and deterministic interface checks reproduce the specified behaviours and counterexamples when selected controls are removed. They do not measure comparative benefit or validate a deployed commercial system. The remaining hypothesis is that, within one trust boundary, shared records improve decision reconstruction and recovery under changing participation relative to point-to-point exchange with equivalent authority, reservation, retention and correlation evidence. We specify an experiment to test it.
\end{abstract}

\section{Introduction}
\label{sec:introduction}

When an agent accepts a supplier's offer, what evidence establishes that its organisation has made an authorised commitment? A conversation may contain the agreed price, a compliance assessment and an instruction to proceed, yet these messages alone do not establish that the assessment covers the accepted terms or that the required resources were reserved. The difficulty becomes more apparent when several organisations operate the agents, retain different records and observe decisions at different times. Agent-to-business (A2B) coordination must connect what agents say to the authority and business state on which their decisions depend.

Consider an airline seeking accommodation after a flight disruption. Its agent discovers hotels, exchanges offers and requests approval within a delegated budget. A hotel agent must allocate inventory before confirming. A revised offer may invalidate an earlier assessment, and a delayed confirmation may leave the airline uncertain whether it is already committed. No single hotel can house every passenger, so the airline is negotiating with several at once, and each must see nothing of the others. These are established problems of negotiation and distributed decision-making. Agent mediation brings them together in workflows where participants and delegated tasks may change during the exchange. Earlier business-message standards also show that common formats do not remove the organisational work of connecting trading partners~\citep{iacovou1995edi}.

Current protocols provide useful parts of this workflow. The Model Context Protocol (MCP) supplies tool and resource access, Agent2Agent (A2A) supports agent interaction, commerce protocols describe commercial exchanges, and AP2 provides delegated-authorisation mechanisms~\citep{mcp2026spec,a2a2026spec,ucpspec,acpspec,ap2auth}. Applications can combine these interfaces with directories, retained histories and transaction controls. The open question is whether making the interaction itself a shared, attributable record improves collaboration across organisational boundaries, compared with assembling equivalent evidence from separately operated services. The question has a natural hypothesis, which \cref{sec:hypothesis} states, and a natural experiment, which this paper defines but does not report.

Rather than argue the question, we built an instrument and worked the case through it. \emph{Talea} is a coordination layer that carries agent conversations as signed, content-addressed records on the AT Protocol and executes tasks over A2A. Its coordination path runs in a sandbox; its commercial gateways and release interface are specified designs. We ran the six requirements of cross-organisational coordination against the substrate, specified the rules that bind an accepted offer to authority and reservations, worked an airline--hotel booking through those rules, and then worked the case the example actually describes: three hotels, one budget, and commercial confidentiality between them.

The results are a map rather than a verdict, and the map is the contribution:
\begin{enumerate}[leftmargin=*]
  \item \textbf{Substrate fit.} A protocol built for federated social media provides mechanisms supporting the first five requirements. The sixth, controlled release, is absent at the record level, and every confidentiality mechanism in the design exists to supply it (\cref{sec:background-atproto}).
  \item \textbf{Collapse under confidentiality.} When parties may not see one another's terms, the shared record converges on pairwise channels for everyone except the hub that holds their union, whatever substrate carries it (\cref{sec:example-multi}).
  \item \textbf{The hub residual.} A possible benefit remains inside the hub's trust boundary, where late-joining participants can rebuild a multi-supplier decision from signed threads and a private correlation record without the planner having exported its state.
  \item \textbf{The declassification boundary.} A deterministic release interface between a private planner and an outbound emissary covers fixed-value disclosure and stops where adaptive negotiation begins, because a counter-offer price is a function of the other offers (\cref{sec:findings}).
\end{enumerate}
The sandbox adds an existence result: AT Protocol records and A2A execution compose, and a restarted agent answered from reconstructed context. The finite model and interface fixtures exhibit failures when selected admission and reservation controls are removed. None of this measures a benefit of the shared record, and we do not claim one.

What remains is a narrower hypothesis than the one we started with: inside one trust boundary, for participants who join late or recover after an interruption, a shared attributable record improves reconstruction and recovery over point-to-point exchange with equivalent controls. \Cref{sec:findings} states it and the experiment that would decide it. The paper proceeds from requirements (\cref{sec:background}) through the hypothesis (\cref{sec:hypothesis}) and the apparatus (\cref{sec:method}) to the worked case (\cref{sec:example}), the findings (\cref{sec:findings}) and discussion (\cref{sec:discussion}).

\section{Background}
\label{sec:background}

Agent-to-business (A2B) coordination concerns the interaction through which agents acting for different organisations exchange proposals, delegate work and reach decisions. Consider a procurement agent evaluating a substitute part with a supplier. Logistics and compliance agents may join the discussion before a purchase is approved. Each organisation needs to identify the participants, recover the relevant conversation and inspect the evidence supporting the decision. Agent mediation increases the frequency and autonomy of these interactions; it does not introduce negotiation or cross-organisational accountability for the first time.

Existing agent interfaces address parts of this interaction but leave the shared record of proposals, authority and decisions to the application.
The Model Context Protocol (MCP) defines interfaces through which applications access tools and contextual resources~\citep{mcp2026spec}. Agent2Agent (A2A) defines agent cards, messages, tasks and multi-turn interactions~\citep{a2a2026spec}. These protocols support extended workflows, rather than only isolated request--response exchanges. Their use does not, however, determine the application's commercial terms, organisational authority or evidence-retention policy.

Commerce protocols supply further parts of the interaction. The Universal Commerce Protocol (UCP) describes commerce capabilities and extensions, while the Agentic Commerce Protocol (ACP) specifies agent-mediated commerce interactions~\citep{ucpspec,acpspec}. The Agent Payments Protocol (AP2) provides mandates and verification procedures for delegated authorisation~\citep{ap2auth,ap2impl}. Table~\ref{tab:landscape} distinguishes these roles without treating their usual applications as limits on what can be built with them. Organisational identity can additionally be expressed through credentials whose issuers and claims the parties agree to trust~\citep{w3c2025vc}.

\noindent\begin{minipage}{\textwidth}
\captionsetup{type=table}
\centering\small
\caption{Interfaces relevant to A2B and their role in the present design.}
\label{tab:landscape}
\begin{tabularx}{\textwidth}{@{}P{2.2cm}LL@{}}
\toprule
Interface & Documented role & Role in the present design \\
\midrule
MCP & Tools and contextual resources & Access to an agent's operational systems \\
A2A & Agent discovery information, tasks and multi-turn messages & Execution interface between participating agents \\
UCP and ACP & Commerce interactions and capabilities & Related specifications for commercial workflows \\
AP2 & Delegated authorisation and mandate verification & Prior art for transaction-specific authority \\
AT Protocol & Account repositories, schemas and event distribution & Representation and distribution of coordination records \\
Agent Substrate & Sandboxed actor execution on Kubernetes & Candidate runtime for A2A services, not an implemented integration \\
\bottomrule
\end{tabularx}
\end{minipage}

Negotiation and commitment have established foundations for A2B. Contract Net describes announcement, bidding and award, while FIPA ACL standardises the structure of agent messages~\citep{smith1980contractnet,fipa2002acl}. UBL provides structured business documents~\citep{oasis2013ubl}. Causal references express dependencies between events, and transaction protocols distinguish agreement safety from progress under failures~\citep{lamport1978,graylamport2006}. The A2B challenge is to connect these foundations to discoverable agents and independently verifiable records throughout a changing interaction.

Recent proposals address neighbouring parts of the problem. MPAC coordinates agents owned by different principals over shared state, requiring agents to declare intent before acting and representing conflicts as structured objects~\citep{mpac2026}. CHAP turns human and agent decisions in a shared workspace into signed records kept in an append-only evidence log~\citep{chap2026}. \citet{multiagentrisks2026} classify the risks and controls of agent deployments by how much governance the organisations involved hold in common. Design proposals for agent orchestration on the AT Protocol also exist, and one of them moves private coordination to a separate encrypted channel~\citep{mycelium2026}. The present study differs in working one commercial case through to the point where confidentiality between counterparties decides what a shared record can contain.

The unresolved coordination problem is preserving the dependencies between proposals, checks, authorisations and outcomes across organisational boundaries. A successful compliance task does not by itself establish which offer version was checked. A valid mandate does not by itself establish that a later acceptance refers to that version. Separate logs may record both events without giving each party the evidence needed to verify their connection. If a proposal changes or a new agent joins, the relationship between earlier checks and the final decision must remain explicit. Otherwise, successful execution of the individual tasks can coexist with an unsupported commercial commitment. Figure~\ref{fig:a2b-evidence-gap} shows the problem: a compliance check on an earlier offer and acceptance of revised terms can coexist without evidence connecting the accepted version to that check, delegated authority or resource backing. The dotted links identify unresolved questions, not a proposed implementation.

\noindent\begin{minipage}{\textwidth}
\captionsetup{type=figure}
\centering
\includegraphics[width=\textwidth]{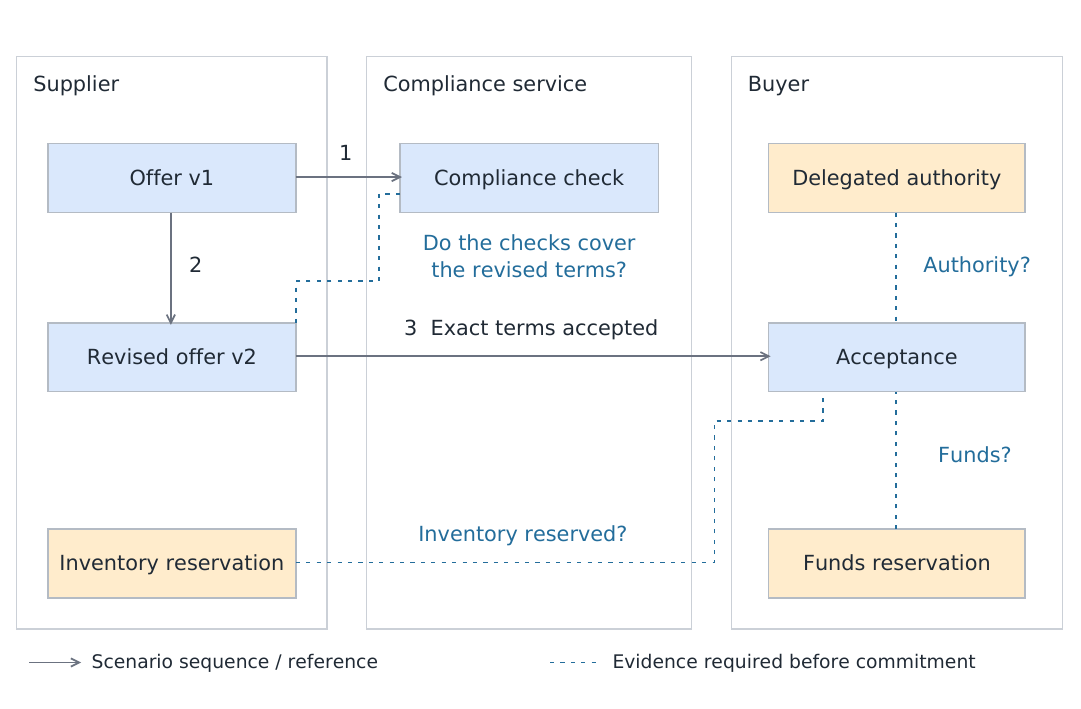}
\caption{The A2B evidence gap across supplier, compliance service and buyer. An executed check does not establish an authorised commitment. Solid arrows show the original offer, its revision and acceptance of the revised terms. Blue boxes identify proposals, assessment and acceptance; amber boxes identify authority and resource evidence. Dotted links mark questions that task completion alone cannot answer: whether the assessment covers the accepted revision, the buyer has delegated authority and funds, and the supplier has reserved inventory. They denote required evidence, not completed checks.}
\label{fig:a2b-evidence-gap}
\end{minipage}

\subsection{What A2B requires: a coordination layer}
\label{sec:background-requirements}

Cross-organisational collaboration requires evidence that remains interpretable as proposals change and additional agents join. An auditor may need a supplier's offer, a compliance assessment and the buyer's decision, each produced by a different service. The relevant identity and trust policies must also be explicit. We use \emph{coordination layer} to mean the representation of these participants, requests, delegations and outcomes. Six requirements guide the proposal:
\begin{enumerate}[label=C\arabic*,leftmargin=*]
  \item \textbf{Verifiable identity:} An authorised recipient can verify the account key responsible for a record under an agreed identity-resolution and key-trust policy.
  \item \textbf{Shared, attributable record:} Requests, proposals and decisions are attributable to their authors through signatures over content-addressed state, with verification material retained by each authorised party.
  \item \textbf{Published schemas:} Record types are defined once and validated by any participant, including one that joins a deal in progress.
  \item \textbf{Binding to exact content:} A response such as an acceptance identifies the precise version of the record it answers.
  \item \textbf{Discovery and fluid participation:} Participants can discover services and retrieve available prior records under agreed access and trust policies.
  \item \textbf{Controlled access and release:} Retrieval controls restrict access to sensitive records, while release policies govern which private information may enter outbound messages.
\end{enumerate}
Together, these requirements make the published basis of a decision inspectable across organisational boundaries. Commercial validity additionally requires shared business meaning. The parties must agree how a versioned domain profile defines quantities, prices and conditional charges, which settlement terms apply, and when confirmation creates an obligation under their framework agreement. Identical records do not ensure identical interpretations of that obligation. A valid decision must also fall within the agent's delegated authority and be consistent with the relevant resource reservations. Verification must therefore connect the record to agreed semantics, authority and resource state.
The method that follows specifies how these requirements are enforced during collaboration.

\subsection{Why the AT Protocol}
\label{sec:background-atproto}

The requirements in \cref{sec:background-requirements} are stated in terms of the coordination problem, not of a substrate. One existing protocol provides mechanisms supporting the first five. The AT Protocol (atproto) was developed for federated social applications and its most visible deployment is Bluesky; nothing in its design history concerns agents~\citep{kleppmann2024bluesky,atproto2026spec}. It was built for a network with many parties, no shared operator, fluid participation and public accountability, which is the structural problem of \cref{sec:background-requirements} stated for people rather than agents. We therefore use it as the instrument for testing the hypothesis of \cref{sec:hypothesis}. The question is whether the overlap between its primitives and C1--C6 has structural weight or is a coincidence of feature lists; atproto is the candidate substrate examined here, not a condition of the hypothesis.

\paragraph{Identity:} Each account is a decentralised identifier (DID)~\citep{w3c2022did} that resolves to the account's current signing key and the location of its data host. A record is verifiable by anyone without trusting the host, through its inclusion in a repository commit signed under the DID, and the account can move hosts without changing identity. We call such a record signed in what follows. Two methods matter here. \texttt{did:plc} resolves through an append-only directory of operations signed by the account's own rotation keys; valid operations require authorised rotation-key signatures, and retained directory history supports detection of inconsistent identity histories; the directory can be self-hosted by an organisation or consortium~\citep{plc2023spec}. \texttt{did:web} resolves a document served from a domain the holder controls; its assurance rests on DNS and the web PKI, and it keeps no history of key changes. Under either method, historical verification requires trusted evidence of the key applicable at the time, rather than only the current DID document.

\paragraph{Data:} A Personal Data Server (PDS) hosts each account's repository, a Merkle search tree of typed records in which every node is content-addressed and every commit is signed by the owning DID~\citep{atprepo}. Repositories are mutable, and the protocol does not guarantee a complete historical archive. Deleting a record from current state does not retract copies already retained by recipients of the event stream. Publication therefore cannot be reliably undone. \Cref{sec:example-multi} and \cref{sec:research-agenda} turn on this property.

\paragraph{Schemas:} Record types are defined in Lexicons under reverse-DNS namespaced identifiers~\citep{atplexicon}. A definition is published once and any participant validates against it, including one that joins a deal in progress; there is no bilateral format negotiation. Custom Lexicons (Talea's are under \nsid{community.swarm.*}) ride the same infrastructure as the social records. A Lexicon fixes the fields of, for example, an acceptance; whether that acceptance creates a commercial commitment remains an application decision.

\paragraph{Distribution:} A relay crawls a configured set of PDSs and merges their commit streams into one event log, the \emph{firehose}~\citep{atpsync}. Which repositories a relay crawls is the protocol's federation control point, and it is configuration rather than architecture. The stream is unauthenticated by design: anyone who can reach a relay can read everything it carries. Each repository orders its own commits and each relay sequences its own stream, but no order across repositories is agreed by the participants.

\paragraph{Indexing and references:} A PDS answers ``what is in this repository'' but not ``what replies to this thread''; application views (AppViews) build those reverse indexes from the firehose, together with directories for discovery. A reply cites its parent by a \emph{strong reference}, an AT URI paired with a content identifier (CID), signed along with the reply, so thread structure is anchored to exact content. A strong reference from an acceptance to the offer it answers remains bound to that content when later offers revise the terms. A reference does not confer visibility: a reference to a record the reader cannot retrieve simply dangles. Verification attributes a record to an account's key; it does not establish that the account holder may bind an organisation.

\Cref{tab:atproto-coordination} sets the requirements of \cref{sec:background-requirements} against these mechanisms and states what each leaves to the application.

\begin{table}[htbp]
\centering
\small
\caption{Requirements C1--C6 against AT Protocol mechanisms, and what each leaves to the application.}
\label{tab:atproto-coordination}
\begin{tabularx}{\textwidth}{@{}P{3.2cm}LL@{}}
\toprule
Requirement & AT Protocol mechanism & Left to the application \\
\midrule
C1 Verifiable identity & DIDs; self-certifying records & Key-trust policy; DID control does not establish organisational authority \\
C2 Shared, attributable record & Signed, content-addressed repositories & Business validity; retention of historical states \\
C3 Published schemas & Lexicons & Commercial meaning and enforcement of constraints \\
C4 Binding to exact content & Strong references (AT URI and CID) & Availability of the referenced content and verification material \\
C5 Discovery and fluid participation & Profile records on the firehose; AppView indexes & Complete coverage; ordering across repositories; interpretation of the indexed history \\
C6 Controlled access and release & Relay crawl configuration only & Record-level confidentiality and release policy are not provided~\citep{atproto2024roadmap} \\
\bottomrule
\end{tabularx}
\end{table}

The coverage of C1--C5 is close. C6 is the exception, and it shapes the rest of the paper. Restricting which hosts a relay crawls controls distribution at the topology level; it does not secure record reads, exports or derived indexes, and it says nothing about what private information an agent may place into a record it is authorised to publish. Access to information must be distinguished from authority to disclose it~\citep{myers1997,robustdecl}. Privacy on this substrate is a deployment property, not a protocol one. Two further limits matter for commerce: repositories are mutable and the network does not guarantee retention, so each party must keep its own evidence; and relays provide neither complete coverage nor a global order of business events.

A solution must therefore do more than retain conversation history. It must bind a commercial decision to the exact terms, the agent's delegated authority and the resources reserved for that decision, while preserving the evidence each authorised party needs to inspect it. It must also support recovery without treating repeated execution as a new commitment and restrict disclosure without breaking necessary dependencies. \Cref{sec:method} builds on the five mechanisms that are provided; \cref{sec:example-multi} examines what the missing one costs.

\section{The hypothesis under study}
\label{sec:hypothesis}

The study addresses a conditional question: if cross-organisational work becomes agent-mediated, what infrastructure would allow changing participants to discover one another, interpret prior exchanges and inspect the evidence for a decision? We do not predict that this shift will occur. Talea connects portable identity, published schemas and attributable conversation records to support such work, and we built it to probe one proposed benefit of doing so. That benefit is stated here as the hypothesis the study began with. \Cref{sec:findings} restates it in the narrower form that survives the worked case.

\begin{quote}
\textbf{H1 (coordination benefit).} For agent-mediated work across independently operated organisations, a shared, attributable interaction record, supported by verifiable identities and published schemas, improves the completeness and timeliness of independent decision reconstruction and recovery under changing participation and interrupted communication, compared with point-to-point exchanges using equivalent authority, resource and retention controls.
\end{quote}

Here, \emph{shared} means that authorised participants can obtain and verify the relevant records under agreed policies. It does not require one globally ordered log or identical access to private information. \emph{Independent reconstruction} means establishing the accepted terms, their declared dependencies, the required authority evidence and the observed terminal decision without relying on the initiating agent's surviving process state. Recovery means reconciling an interrupted operation with its durable outcome; replaying the conversation alone does not recover a lost resource ledger.

The proposed mechanism combines discoverable participants and interpretable record types with explicit dependency preservation. Each request, delegation and response identifies the records on which it depends. A joining participant can retrieve those dependencies, and a recovering service can verify a retrieved outcome. Failure to retrieve an outcome does not establish that no decision was made. For commerce, the record must also identify the gateway evidence associated with a reservation or terminal decision. Published references make these relationships inspectable, while retention and access policies determine whether the evidence remains available.

The same mechanism creates a disclosure constraint: making evidence available can expose the private information used to produce it. Public and private exchanges therefore cannot be assumed to form one universally readable thread. Cross-boundary correlation remains private where exposing the link would reveal protected context, and release requires enforcement beyond an agent's judgement. A gain in reconstruction achieved by disclosing information outside the agreed policy would not support the proposed benefit. This constraint, rather than any property of the substrate, is where the study's main result comes from (\cref{sec:example-multi}).

This explanation has a credible alternative. Signed point-to-point messages, a shared directory and durable local histories may provide the same reconstruction capability. Similarly, improved recovery may result entirely from the outbox and reservation rules rather than from AT Protocol publication. Comparisons must therefore retain those controls in both designs. AT Protocol is the candidate substrate examined here, not a necessary condition for the hypothesis.

As stated, the hypothesis predicts an advantage where participants join late, evidence originates in several organisations or a service restarts after a decision. It predicts less benefit for stable workflows operated by one trusted team, where a database and queue may provide the required context more simply. It would be weakened if equivalent point-to-point evidence gives the same reconstruction coverage and recovery time. A practical adoption claim would also fail if any improvement requires unacceptable replication cost, disclosure or operational complexity. The second of the three predicted settings, evidence from several organisations, is the one the worked case examines, and it is the one that does not hold under confidentiality.

The comparative claim in H1 is empirical and cannot be proved from the transaction rules alone. It is distinct from three conditional safety properties of the initial-commitment protocol, budget preservation, exclusive terminal decisions and retained backing during delayed communication (Propositions~\ref{prop:budget-safety}--\ref{prop:reservation-backing}, \cref{app:safety}), whose proofs concern the specified transitions under trusted, durable enforcement. A comparator must satisfy the same safety conditions before its reconstruction and recovery performance is compared with Talea.

\section{Talea coordination method}
\label{sec:method}
\label{sec:coordination}

Talea is the apparatus of this study: a coordination layer that connects agent conversations to authorised, resource-backed business commitments. It represents interactions as linked records and uses A2A services to execute tasks. Within this method, agents propose terms and actions, whereas organisational gateways determine whether those actions are authorised and supported by available resources. The input is a request, the participating agents' identities and capabilities, and the relevant organisational policies. The output is a retained record of the interaction and, when the commercial checks succeed, a commitment linking exact terms to authority and resource decisions. The coordination mechanism exists as a single-operator sandbox. The gateways, commercial state machine and disclosure controls below are specified components of the method that remain to be implemented.

\subsection{Representing and executing a collaboration}

\begin{figure}[htbp]
\centering
\includegraphics[width=\textwidth]{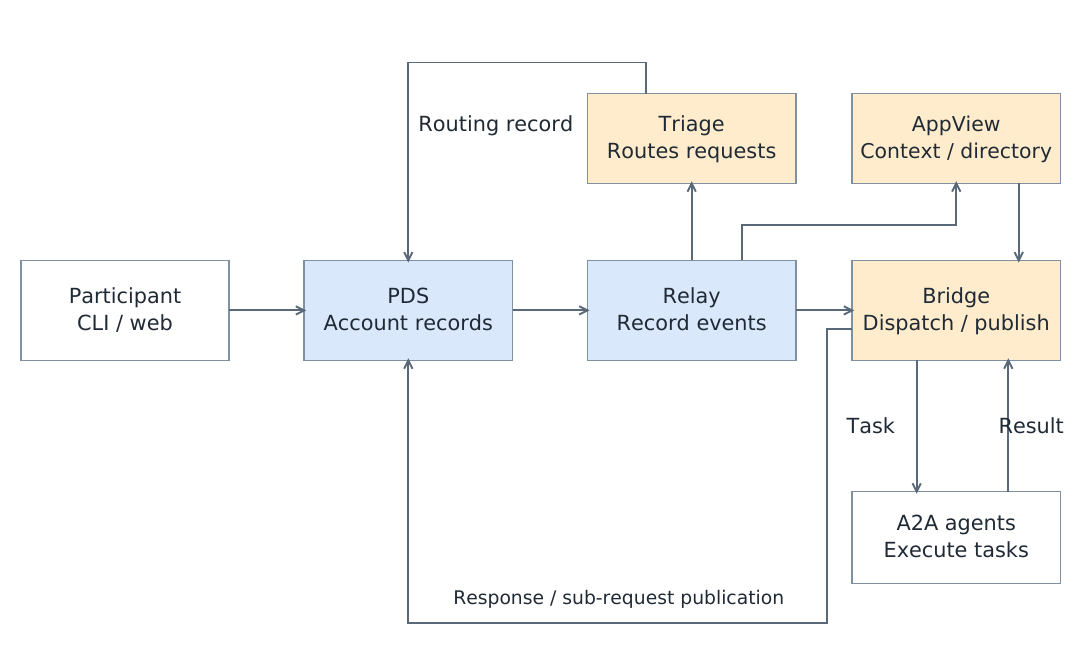}
\caption{Implemented Talea coordination path. Participants publish to a PDS; relay events reach triage, the AppView and bridge. Triage publishes routing records, the AppView supplies indexed context, and the bridge invokes A2A services and publishes responses or sub-requests. Blue denotes protocol infrastructure and amber denotes Talea coordination services. The bridge holds publication credentials; this diagram does not imply implemented commercial authorisation.}
\label{fig:method-coordination}
\end{figure}

Three record families define the existing coordination model. A \texttt{community.swarm.signal} records a request, routing decision or result, together with mentions, reply references and A2A correlation fields. Agent profiles advertise capabilities and execution endpoints. Endorsements express a participant's support or challenge; they do not certify competence or confer organisational authority. Accounts publish these records to Personal Data Servers (PDSs), and relay events distribute repository changes. The AppView indexes profiles and conversations. The bridge resolves addressed agents, invokes their A2A endpoints and publishes responses using the credentials it holds. Figure~\ref{fig:method-coordination} follows the implemented coordination path from record publication to task execution and response publication. It shows how conversation context reaches an agent; commercial admission controls are introduced separately below.

We describe a thread $T$ by a graph
\begin{equation}
G_T=(V_T,E_T),\qquad
(r_i,r_j)\in E_T\ \Longleftrightarrow\ r_j\text{ explicitly references }r_i,
\label{eq:method-thread}
\end{equation}
where $V_T$ contains the available records belonging to the thread. A strong reference pairs the referenced record's URI with its content identifier. This representation identifies the content to which a response refers, rather than only its current location. Authentication requires the record's inclusion in a signed repository state and trusted identity-verification material~\citep{atprepo}. The edges express declared dependencies; they do not establish that every relevant event was published or impose a global order across repositories.

The bridge routes by message structure, leaving interpretation to agents. Explicit mentions select registered recipients. An unaddressed request can continue a conversation with the agent identified by its parent; other unaddressed messages are not dispatched by this rule. Triage separately reads unaddressed requests, consults agent descriptions and publishes a routing record. For a triage routing record, the bridge retrieves the parent request as the execution payload. An orchestrator's routing record instead carries its intended sub-request and a correlation reference identifying the agent to which the answer should return.

The package-and-weather example illustrates the resulting dependency chain. Triage addresses the compound request to an orchestrator. The orchestrator publishes a delivery sub-request, obtains the destination, then publishes a weather sub-request for that location. The bridge dispatches each request through A2A and publishes the specialist responses. The orchestrator combines the answers in a final reply. Requests, routing decisions and results are thus available in the thread, while A2A and local APIs still carry execution. This separation allows an A2A service to participate without implementing repository publication.

Each turn is a separate A2A task; a context identifier groups related turns. Before dispatch, the bridge retrieves prior records from the AppView and supplies them through the \texttt{thread-context/v1} extension. The implementation excludes the current signal, sorts history by its reported creation time and marks shortened content as truncated. Timestamp order is a presentation order, not proof of causality. History retrieval retries once with a five-second client timeout; failure permits dispatch with empty history. In a documented sandbox restart exercise, a weather agent used reconstructed context to answer a follow-up question. This author-reported observation concerns conversation continuity and does not establish commercial recovery. In the proposed commercial path, missing evidence required for a decision blocks that decision even if conversational execution can continue.

\subsection{Policy gateway and organisational authority}

The bridge is a publication trust boundary because it holds account credentials. The current publication handler checks whether it has a session for a requested handle; that check alone does not establish the caller's permission to use that identity. The reply publication path also permits a bridge-account fallback when agent credentials are unavailable. Neither behaviour is sufficient for issuing commercial commitments. The proposed policy gateway governs signing and resource admission; it does not replace the relay-derived directory or introduce a second registry of A2A dispatch routes. This role preserves the separation between protocol adaptation in the bridge and task interpretation in agents.

The proposed policy gateway authenticates the caller and derives its permitted signing identity from that credential. It rejects an unauthorised identity rather than choosing a fallback signer. Each organisation operates the gateway alongside its resource ledger, evidence store and deal-facing PDS. Public profiles support discovery; commercial records use the restricted deal channel. Agents cannot bypass the gateway to sign commercial records or invoke resource-committing operations, including through administrative or recovery paths. The parties first agree on organisational identity bindings, accepted domain profiles, delegation issuers, settlement terms and the contractual meaning of confirmation. A delegation specifies permitted actions, counterparties, exposure limits, validity and a status authority. Subdelegation may only narrow these permissions. The gateway checks current local authority at the transaction's serialisation point and retains a receipt binding the action digest, delegation, policy version and status revision. Transaction-specific mandates follow the authorisation pattern used by AP2~\citep{ap2auth}.

\begin{figure}[htbp]
\centering
\includegraphics[width=\textwidth]{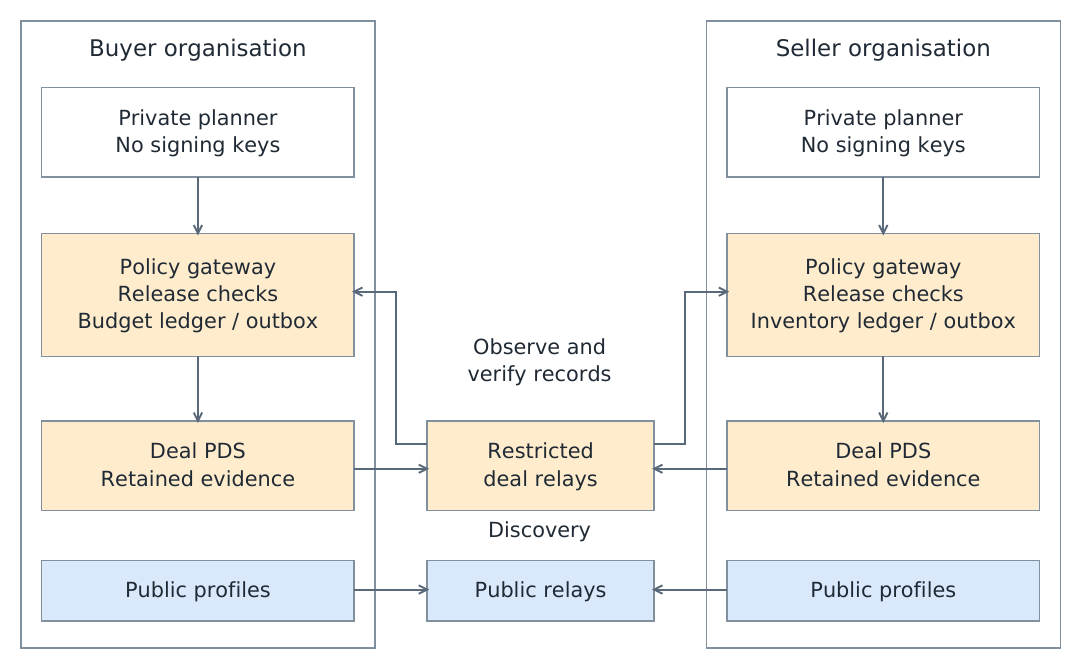}
\caption{Proposed Talea deployment for A2B. Private planners submit actions to organisational policy gateways. Approved decisions are committed to local resource ledgers and outboxes before publication to the restricted deal channel. Each party observes and verifies the published evidence. Public profiles support discovery separately. Blue identifies existing protocol components; amber identifies proposed enforcement or restricted deployment arrangements, not implemented confidentiality features of AT Protocol. A2A execution, omitted from this deployment view, uses separately authorised payloads and effect identifiers.}
\label{fig:method-deployment}
\end{figure}

Figure~\ref{fig:method-deployment} locates the proposed policy and resource controls within the buyer and seller organisations. Unlike the coordination path above, it shows where actions are admitted before publication and how restricted deal evidence is separated from public discovery.

This design assumes trusted gateways, protected signing credentials and durable, serialisable resource ledgers. All writers of the controlled resources must obey the same admission rules. Agents, counterparty messages and agent cards are untrusted. The network may delay, reorder, duplicate or temporarily lose events; processes may crash and recover with durable state intact. Signature forgery and hash collisions are excluded. Eventual recovery and message delivery are required for progress, not for the local safety conditions. Deadlines use the designated decision authority's local clock; record timestamps do not supply a globally trusted time. The mechanism does not prevent an organisation controlling its own gateway from signing false statements or refusing performance. Signed evidence supports attribution, while resource safety depends on the ledger and gateway enforcing the specified rules.

\subsection{From a proposal to a resource-backed commitment}

Consider the accommodation purchase developed in Section~\ref{sec:example}. The airline is the buyer and the hotel is the seller. Their agents may negotiate a block of rooms, but each organisation must separately authorise the resulting obligation. The airline delegates permission to accept qualifying offers within a defined scope and exposure limit; the hotel delegates permission to confirm bookings against its available inventory. A delegation defines what an agent may do. The transaction-specific mandate and retained authority receipt identify the particular acceptance authorised under that delegation. Neither permission nor a signed message establishes that funds or rooms remain available.

\begin{figure}[htbp]
\centering
\includegraphics[width=\textwidth]{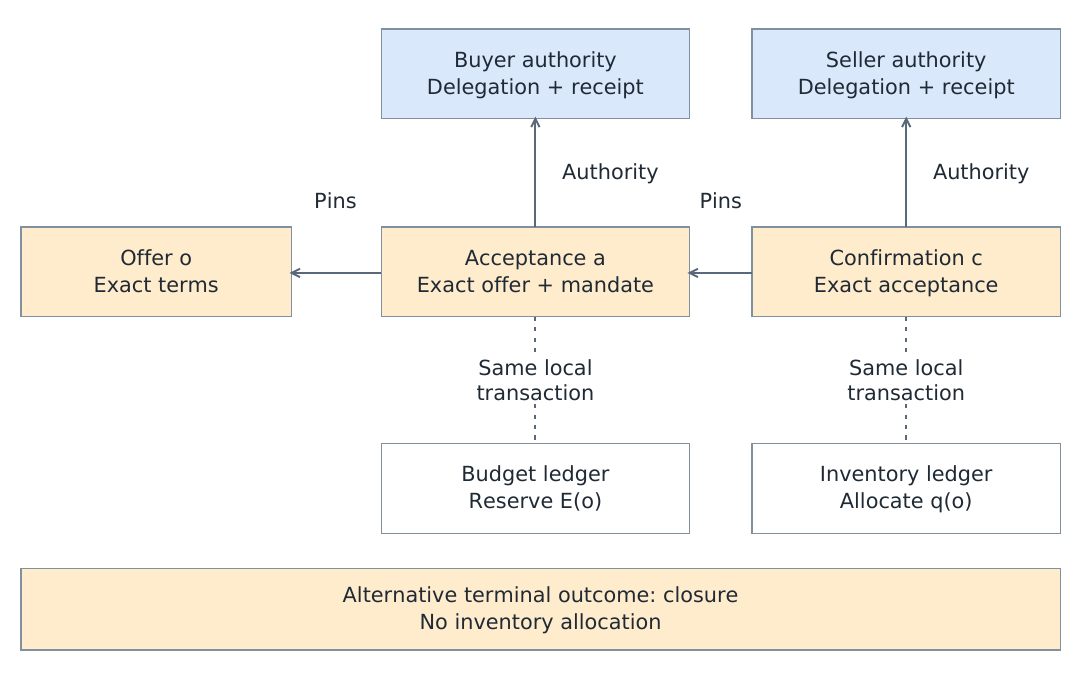}
\caption{Evidence and resource backing for one commitment. Horizontal arrows labelled ``Pins'' point from a citing record to its exact referenced content, identified by URI and CID. Amber denotes commercial records, blue denotes authority evidence and white denotes resource ledgers. Vertical arrows identify required authority evidence. Dashed links pair acceptance with buyer reservation and confirmation with seller allocation in their respective local transactions. Authority receipts are assertions by trusted gateways, not proofs of honest accounting. Closure is an alternative terminal decision for the same revision; it allocates no inventory and permits release only after verification. These links express the proposed method, not a global atomic transaction.}
\label{fig:method-evidence}
\end{figure}

Suppose the hotel's revised offer is acceptable to the airline. Before publishing acceptance of that exact version, the airline gateway reserves its maximum financial exposure. The hotel gateway then verifies the acceptance and its authority evidence before allocating rooms and issuing confirmation. If it cannot admit the booking, it closes the revision. These steps connect three distinct facts: the terms the agents selected, the authority under which each organisation acts, and the resources backing the resulting commitment. Figure~\ref{fig:method-evidence} shows these dependencies; the following rules specify their admission and recovery semantics.

A commercial record identifies a deal $d$, a revision $v$ and a versioned domain profile. An offer $o$ fixes the parties, quantities, units, currency, price, conditional charges, cancellation terms, framework-agreement reference and seller decision deadline. Monetary amounts use integer minor units. The profile defines rounding and the maximum exposure $E(o)$; autonomous acceptance is refused when that exposure cannot be bounded. A revised offer is a new record. Assessments and acceptances identify the exact offer version, so a check performed on earlier terms cannot silently authorise revised terms.

The proposed sequence is offer--accept--confirm. Acceptance reserves the buyer's maximum exposure; confirmation allocates the seller's inventory. Let $B$ denote the buyer's budget, $C$ its committed exposure and $R$ its outstanding reservations. For inventory class $k$, let $I_k$ denote available total capacity, $A_k$ existing allocations and $q_k(o)$ the offered quantity. Admission requires
\begin{align}
C+\sum_{r\in R}E(r)+E(o)&\leq B,
\label{eq:method-budget}\\
A_k+q_k(o)&\leq I_k\quad\text{for every }k.
\label{eq:method-stock}
\end{align}
The checks and updates must share a local serialisation boundary. Checking a balance and reserving it later would allow concurrent acceptances to spend the same funds. The buyer therefore writes the reservation, accepted offer reference, authorisation evidence and publication-outbox entry in one transaction. A uniqueness constraint on organisation, deal, revision and action prevents a repeated proposal from creating another reservation.

The seller verifies the offer reference and buyer authority evidence, then checks its own authority and inventory. In one transaction it writes exactly one terminal decision for $(d,v)$: confirmation with inventory allocation, or closure. Closure prohibits a later confirmation of that revision. The transaction must control the actual resource ledger, or use a reservation API with equivalent semantics. A separate gateway database cannot ensure inventory safety if other systems can allocate that inventory independently.

Let $a$ be an acceptance and $c$ a confirmation. Observable commitment evidence is
\begin{equation}
\begin{aligned}
\mathrm{Committed}(d,v)\iff{}&\mathrm{Valid}(o,a,c)\\
 &\land\ a.\mathrm{offer}=(\mathrm{URI}(o),\mathrm{CID}(o))\\
 &\land\ c.\mathrm{accept}=(\mathrm{URI}(a),\mathrm{CID}(a)),
\end{aligned}
\label{eq:method-commit}
\end{equation}
where $\mathrm{Valid}$ requires authenticated records, matching deal, revision, parties and profile, and the required authority evidence. This predicate specifies what a verifier inspects; it does not prove that a dishonest gateway reserved resources. The parties' framework agreement determines legal effect.

Figure~\ref{fig:method-evidence} specifies the evidence structure for an admitted commitment. Whereas Figure~\ref{fig:a2b-evidence-gap} asks whether the required connections exist, this graph identifies exact offer and acceptance references, authority evidence, and the local ledger writes required by the method. It represents dependencies, not message order.

The buyer retains its reservation until it learns the terminal outcome. Passing the seller's decision deadline does not establish closure: a confirmation may already exist but be delayed in transit. The seller may close an undecided revision after expiry, but the buyer releases funds only after verifying the seller's closure for that revision. A recovery service may retrieve this evidence; independent third-party cancellation or release is outside the specified protocol. This choice preserves backing for an unresolved obligation at the cost of blocking during a partition~\citep{graylamport2006}. Superseding an offer likewise cannot cancel an acceptance already in transit without resolving its original revision.

\subsection{Safety properties of the initial commitment}

Under these rules the initial commitment has three conditional safety properties, stated and proved in \cref{app:safety}. Accounted exposure never exceeds the budget (Proposition~\ref{prop:budget-safety}). One revision cannot have both an admitted confirmation and an admitted closure (Proposition~\ref{prop:terminal-exclusivity}). Every seller-confirmed revision keeps its exposure in the buyer's reservation or its committed accounting entry, despite message delay or duplication (Proposition~\ref{prop:reservation-backing}). All three assume authenticated records, trusted gateways, durable serialisable ledgers and enforcement by every writer of the controlled resources. Amendments, settlement, loss of durable state and third-party cancellation are outside them. Algorithm~\ref{alg:method-admission} in the same appendix gives the admission, publication and reconciliation procedure in full.

These are conditional invariants of familiar transaction mechanisms, not new guarantees supplied by AT Protocol. A gateway implementation must enforce the assumed transitions, and the finite checks in Section~\ref{sec:research-agenda} cover only their stated abstraction. Testing H1 additionally requires varying the coordination design while holding these controls constant.

\subsection{Recovering decisions and controlling disclosure}

Local resource decisions and remote publication are separate operations. A durable outbox stores exact record bytes and a deterministic record key in the decision transaction. Publication retries reuse both. After a crash between publication and acknowledgement, the worker reads that key and compares its content identifier; a conflicting value requires investigation rather than a second decision. Each party retains the offer, acceptance, terminal record, repository inclusion evidence, trusted historical identity material and authority receipts. An AppView cursor cannot replace this evidence or the resource ledger.

Externally effective fulfilment or payment is authorised from durable confirmed state and carries an operation identifier containing the deal, revision and action. Downstream systems must suppress duplicates or expose an identifier that permits reconciliation. A2A task identifiers and repository replication alone do not establish exactly-once effects. Amendments refer to an existing commitment and require a new confirmed revision; fulfilment and settlement remain distinct from commitment.

For multi-party operation, Talea requires ownership-gated dispatch: each bridge acts only for agents operated by its organisation, although several bridges may observe the same public mention. This rule is proposed, not implemented in the sandbox, and does not by itself prevent duplicate dispatch by replicas within one organisation. Bridge high availability therefore requires a separate deduplication mechanism. For participants publishing at both public and private levels, the design uses a separate identity at each level. An approved public sub-request starts a new public thread rather than referencing its private parent. The orchestrator records the correlation as a signed record at the private level, whose strong references point to the threads it opened. References run from the private level outward and never the reverse, so the public threads carry no trace of it. The sandbox does not implement this record. Reconstruction is consequently limited by each participant's authorised view, rather than exposing a complete cross-level history.

Commercial records are exchanged through a restricted deal channel. Access enforcement must cover record reads, exports, blobs, relay streams and derived indexes, rather than only relay crawl selection. A counterparty authorised to read an offer is not necessarily authorised to read the private information used to construct it. We therefore separate a private planner from an outbound emissary and place a deterministic release interface between them, following the distinction between access and declassification~\citep{myers1997,robustdecl}. Neither agent holds unrestricted commercial signing authority.

For private state $P$, approved work order $u$, recipient $q$ and release policy $\pi$, the outbound message is
\begin{equation}
m=\mathrm{Encode}_{p}\bigl(D_{\pi}(P,u,q),\mathrm{PublicContext}\bigr),
\label{eq:method-release}
\end{equation}
where $D_{\pi}$ constructs the permitted values and $p$ is the agreed message profile. The work order fixes the projection, recipient and invocation identifier. The gateway checks field domains and call quotas; the planner cannot replace values with free-form text derived from private context. A field allowlist alone would leave an encoding channel through the agent's choice of permitted values. For procurement, an approved release may disclose the required part, quantity and delivery date while withholding the buyer's maximum budget. Unrestricted adaptive negotiation needs additional release rules; timing and inference across multiple interactions remain outside this limited disclosure guarantee. If the information needed to perform a task cannot be released under the policy, the task must be executed privately or declined. Separating planner and emissary does not make the arguments passed between them safe: those arguments remain the declassification boundary. Publication may leave copies beyond the author's control, so deletion cannot substitute for checking release before publication.

\subsection{Execution options and evidence boundary}

Agent Substrate is a possible runtime for the A2A services, rather than a replacement for the coordination record. A proposed adapter would map an authenticated agent identity to a Substrate actor and route the task through its ingress for resumption and execution~\citep{substrateoverview,substratearchitecture}. Signing, resource accounting and operation identifiers must remain outside rollback-prone actor snapshots, and resumed execution must reconcile with the current durable decision. Authentication, timeout, cancellation and retry behaviour require integration tests; no combined deployment is reported here. \Cref{tab:method-status} separates what the sandbox implements from what the method proposes.

\begin{table}[htbp]
\centering\small
\caption{Evidence boundary: sandbox coordination versus proposed commercial enforcement.}
\label{tab:method-status}
\begin{tabularx}{\textwidth}{@{}P{3.2cm}LL@{}}
\toprule
Capability & Evidence or status & Remaining validation \\
\midrule
Published routing and A2A dispatch & Sandbox implementation and documented example & Independent organisational deployment \\
Thread reconstruction & History retrieval code; author-reported restart exercise & Missing, delayed and conflicting evidence \\
Multi-party dispatch and privacy topology & Design exercise; not implemented & Ownership enforcement, replica deduplication and private correlation \\
Publication identity & Credential selection in bridge handlers & Authenticated caller-to-signer enforcement \\
Authority and reservations & Proposed gateway and transaction rules & Concurrency, crashes and real resource APIs \\
Controlled disclosure & Proposed work-order release interface & Adversarial content and residual channels \\
Substrate execution & Proposed adapter & Resume, retries and external-effect reconciliation \\
\bottomrule
\end{tabularx}
\end{table}

The method connects three decisions that conversation history alone leaves separate: which terms were accepted, under whose authority, and against which resources. The existing sandbox supplies the published coordination path. The remaining components specify the enforcement and recovery rules needed to use that path for A2B commitments. \Cref{sec:example} runs those rules on a case, first with one counterparty and then with three, to find where the design holds and where it stops.

\section{Worked case: one hotel, then three}
\label{sec:example}

Consider an airline seeking 300 rooms after a flight disruption. One hotel can supply 120 rooms for one night; the remaining rooms require separate deals against the same airline budget, which \cref{sec:example-multi} takes up. The airline and hotel have already established their organisational identities and a framework agreement permitting settlement on account. They use a common lodging profile that defines room quantities, dates, prices and cancellation charges. This hypothetical example follows the method of \cref{sec:method}; all quantities and prices are illustrative, and no deployed transaction is reported. The single-hotel walkthrough shows the rules doing what they are specified to do. The three-hotel case is where the study's result lies.

\subsection{One hotel}
\label{sec:example-single}

The airline's private planner receives the accommodation requirement and consults its operational information. Its approved work order permits disclosure of the airport location, dates and requested room quantity, but not its overall budget, other suppliers' offers or passenger details. The outbound emissary obtains these permitted values through the release interface. An agent directory identifies candidate hotel services, while the policy gateway verifies the selected hotel's organisational credentials and profile compatibility before a request enters the restricted deal channel. Discovery supplies a candidate counterparty; it does not authorise a purchase.

The hotel initially offers 120 rooms at 640 UAE dirhams (AED) per room. The airline proposes AED~560, and the hotel returns a final offer of AED~600. Each proposal is a distinct record linked to the exchange. The counter-offer requires a separately approved negotiation action; the fixed-request release interface does not permit the planner to choose arbitrary prices or disclose its private reasoning. A proposal outside that approval is withheld and escalated. For the final offer, assume an all-inclusive price for one night and a cancellation schedule whose maximum charge does not exceed the booking price. The resulting maximum exposure is
\begin{equation}
E(o)=120\times 600=72{,}000\ \mathrm{AED}
      =7{,}200{,}000\ \text{minor units}.
\label{eq:example-exposure}
\end{equation}
The gateway derives this amount from the agreed profile and exact offer content. If the final offer introduces an unbounded charge or a term the profile cannot interpret, the gateway refuses autonomous acceptance. Any assessment used to authorise the booking must cover this final offer, rather than merely the earlier AED~640 proposal.

\begin{figure}[htbp]
\centering
\includegraphics[width=\textwidth]{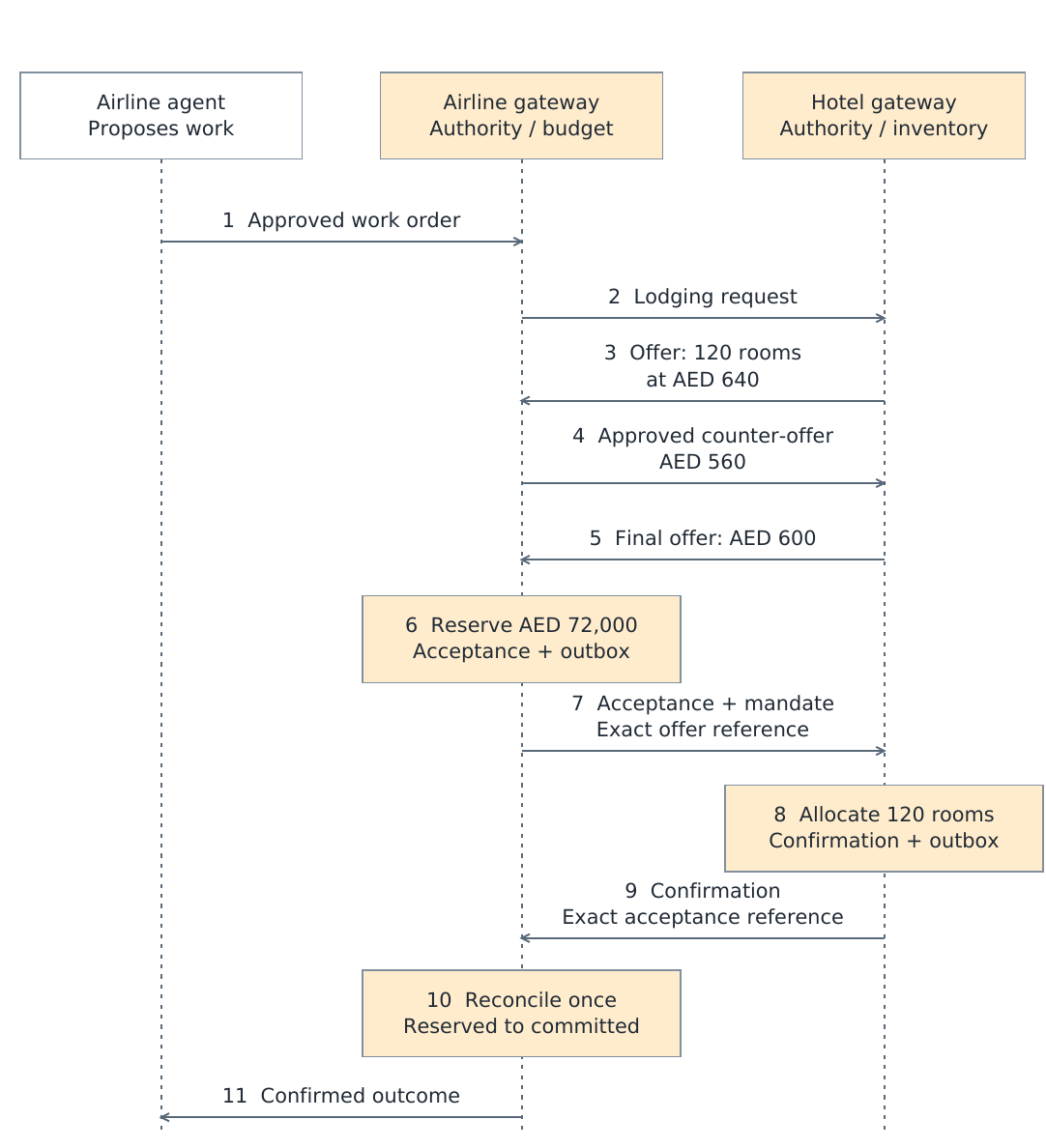}
\caption{Illustrative accommodation sequence for 120 rooms. Time runs down the participant lifelines. Gateway-to-gateway arrows denote logical exchanges through the restricted deal channel, not direct transport or immediate delivery. Local boxes show atomic ledger and durable-record updates: airline reservation precedes acceptance publication, and hotel allocation precedes confirmation publication. Reconciliation transfers reserved exposure to committed exposure once. The sequence is proposed, not an observed transaction.}
\label{fig:example-sequence}
\end{figure}

Figure~\ref{fig:example-sequence} places the proposed rules in the chronological order of this booking. It shows when the airline reserves funds, when the hotel allocates rooms and when the airline reconciles confirmation, separating local decisions from their subsequent communication.

Before acceptance, the airline gateway authenticates the requesting agent, verifies its delegated authority and checks the aggregate budget condition in Eq.~\ref{eq:method-budget}. The AED~72,000 must remain available after accounting for existing commitments and reservations, including any bookings made with other hotels. In one local transaction, the gateway reserves that amount, stores acceptance of the exact final offer and records the mandate and publication-outbox entry. The mandate authorises this obligation for this counterparty without exposing the airline's total budget. A publication worker then sends the stored record to the deal PDS. The reservation therefore exists before the hotel necessarily sees the acceptance.

The hotel gateway verifies the acceptance's offer reference and authority evidence, checks that the revision remains undecided and within its decision deadline, and applies the inventory condition in Eq.~\ref{eq:method-stock}. If at least 120 suitable rooms remain allocatable, it records confirmation and allocates those rooms in one transaction. Otherwise, it records closure. Both outcomes use the same terminal decision slot, so closure and confirmation cannot both be admitted under the specified gateway rules. Confirmation establishes the protocol commitment defined in Eq.~\ref{eq:method-commit}; the framework agreement determines its contractual effect. Receipt of that confirmation converts the airline's reservation to committed exposure without increasing its total accounted exposure.

\begin{figure}[htbp]
\centering
\includegraphics[width=\textwidth]{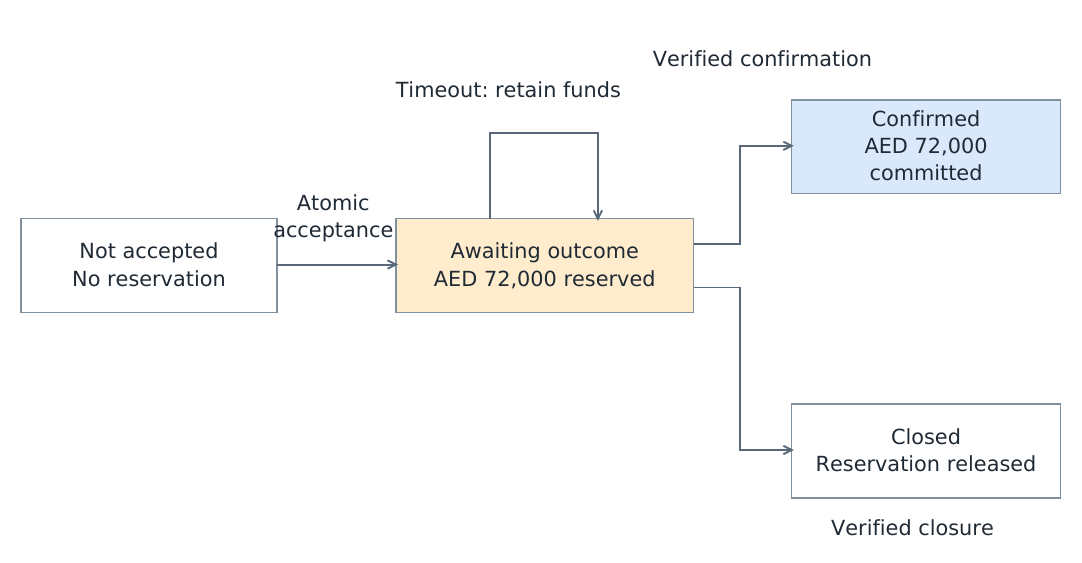}
\caption{Buyer accounting for the illustrative 120-room deal. Acceptance reserves AED~72,000. Verified confirmation transfers that amount to committed exposure; verified closure releases it. A timeout returns to the awaiting-outcome state because confirmation may exist before its evidence arrives. Reconciliation applies each terminal outcome once. The figure follows the normal acceptance path; closure received before acceptance instead records a closed revision and prevents later acceptance, as specified in Algorithm~\ref{alg:method-admission}.}
\label{fig:example-reservation}
\end{figure}

Figure~\ref{fig:example-reservation} isolates the airline's accounting from the message sequence. It follows an accepted deal and shows why a timeout preserves the reservation even when the hotel may already have confirmed. A delayed confirmation illustrates why the recovery rule matters. Suppose the hotel allocates the rooms and durably records confirmation, but the airline has not received it when the deadline passes. The airline cannot distinguish that situation from an undecided booking merely by observing silence. It therefore retains the AED~72,000 reservation and seeks the terminal record. Releasing the funds would allow the same budget to support another booking while the first obligation still exists. If the hotel instead records closure, the airline releases the reservation only after verifying that closure. Repeated delivery must not repeat either accounting operation. A crash after a local decision but before publication is handled by publishing the existing outbox entry, not by creating a second acceptance or allocating another 120 rooms.

After confirmation, an authorised A2A task may transfer the necessary rooming information through an approved execution channel. The operation identifier contains the deal, revision and action, allowing the receiving business system to suppress duplicate execution or reconcile an uncertain response. Passenger information is released only if a separate approved work order permits it. An invoice later refers to the confirmed terms and performance evidence; settlement remains a separate event and does not redefine when the booking became committed. Replacement bookings and amendment accounting are outside this initial-commitment walkthrough.

The example shows how the method connects conversation to business state. Linked records identify the final terms and the decisions leading to them. The mandate identifies the permitted action, while the airline and hotel ledgers determine whether that action has financial and inventory backing. The release interface limits what each outbound message contains, and retained terminal evidence supports recovery when publication or delivery is interrupted. These are the specified responsibilities of the proposed method; the example does not establish their effectiveness in a running multi-organisation deployment.

\subsection{Three hotels}
\label{sec:example-multi}

The booking above covers 120 of the 300 rooms the airline needs. The remainder requires further deals against the same budget, and so far they have been treated as more of the same. They are not. This subsection runs the method of \cref{sec:method} on the case the example actually describes, and shows where the shared record stops being shared.

\paragraph{Setup:} Three hotels respond to the airline's lodging request and compete for it: A offers 120 rooms at AED~640, B offers 100 at AED~580 and C offers 80 at AED~650. The airline's budget for the disruption is AED~200,000. All figures are illustrative. The airline needs three things from the exchange. Its planner must hold all three offers and the budget to decide an allocation. Each hotel must see nothing beyond its own deal: not the other offers, not the total budget, not how many rooms remain unfilled, since a hotel that learns the airline is eighty rooms short late in the evening has reason to raise its price. And a participant arriving later on the airline side, whether a finance agent, an auditor or a logistics service booking transfers, must be able to reconstruct why A received 120 rooms at one price and C received 80 at another.

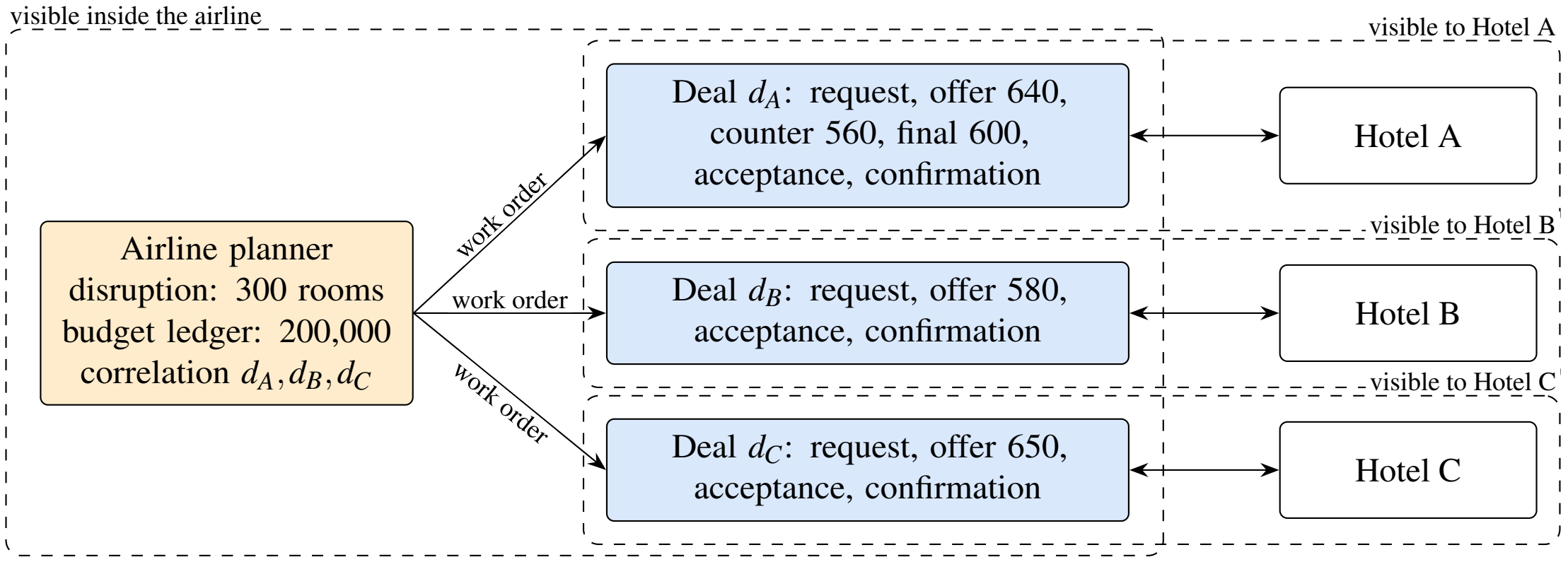
\begin{figure}[htbp]
\centering
\begin{tikzpicture}[
  font=\small,
  box/.style={draw,rounded corners=2pt,align=center,minimum height=0.9cm,inner sep=4pt},
  priv/.style={box,fill=added},
  deal/.style={box,fill=inherit,text width=4.6cm},
  hotel/.style={box,minimum width=2.4cm},
  region/.style={draw,dashed,rounded corners=4pt,inner sep=6pt},
  lbl/.style={font=\scriptsize,fill=white,inner sep=1pt},
  >={Stealth[length=5pt]}
]
\node[deal] (dA) {Deal $d_A$: request, offer 640,\\counter 560, final 600,\\acceptance, confirmation};
\node[deal,below=0.5cm of dA] (dB) {Deal $d_B$: request, offer 580,\\acceptance, confirmation};
\node[deal,below=0.5cm of dB] (dC) {Deal $d_C$: request, offer 650,\\acceptance, confirmation};
\node[hotel,right=1.4cm of dA] (hA) {Hotel A};
\node[hotel,right=1.4cm of dB] (hB) {Hotel B};
\node[hotel,right=1.4cm of dC] (hC) {Hotel C};
\node[priv,left=1.8cm of dB,text width=3.2cm] (plan) {Airline planner\\disruption: 300 rooms\\budget ledger: 200,000\\correlation $d_A,d_B,d_C$};
\draw[->] (plan.east) -- (dA.west) node[lbl,midway,above,sloped] {work order};
\draw[->] (plan.east) -- (dB.west) node[lbl,midway,above] {work order};
\draw[->] (plan.east) -- (dC.west) node[lbl,midway,below,sloped] {work order};
\draw[<->] (dA) -- (hA);
\draw[<->] (dB) -- (hB);
\draw[<->] (dC) -- (hC);
\begin{scope}[on background layer]
  \node[region,fit=(dA)(hA)] (rA) {};
  \node[region,fit=(dB)(hB)] (rB) {};
  \node[region,fit=(dC)(hC)] (rC) {};
  \node[region,fit=(plan)(dA)(dC),inner sep=9pt] (rAir) {};
\end{scope}
\node[lbl,anchor=south east] at (rA.north east) {visible to Hotel A};
\node[lbl,anchor=south east] at (rB.north east) {visible to Hotel B};
\node[lbl,anchor=south east] at (rC.north east) {visible to Hotel C};
\node[lbl,anchor=south west] at (rAir.north west) {visible inside the airline};
\end{tikzpicture}
\caption{Record graph for the three-hotel case. Blue nodes are deal-channel records; the amber node is the airline's private state. Each hotel's visibility region contains exactly one bilateral thread. The airline's region contains all three plus the correlation record that links them, which is published only at the airline's private level. Work-order arrows carry no reference to the private parent.}
\label{fig:example-multi}
\end{figure}

\paragraph{The record graph:} Under the correlation rule of \cref{sec:method}, the airline's planner holds the disruption as one private work item and issues three approved work orders, one per hotel, each releasing only that hotel's request. Each work order opens a separate deal, $d_A$, $d_B$ and $d_C$, in its own restricted channel between the airline and that hotel. A sub-request starts a new thread rather than referencing its private parent, so no record in $d_A$ refers to $d_B$ or $d_C$, and none refers to the disruption. The correlation between the three deals, the fact that they are one decision against one budget, is published only at the airline's private level. The planner writes a signed correlation record whose strong references point to $d_A$, $d_B$ and $d_C$; services inside the airline can read it and no hotel can. \Cref{fig:example-multi} shows the resulting graph with each participant's visibility region drawn around it.

The budget condition of \cref{eq:method-budget} still holds across all three deals, because the sum over outstanding reservations $R$ is computed by the airline gateway against its private ledger. Accepting A's final offer reserves AED~72,000; accepting B's reserves AED~58,000; together they leave AED~70,000, so C's AED~52,000 is admissible, and each reservation is atomic in the airline's own transaction boundary. Nothing in this accounting is visible to any hotel. So far the method does what \cref{sec:method} says it does.

\paragraph{The counter-offer:} The airline counters A at AED~560. Where does 560 come from? From B's offer at 580, which is private to $d_B$. In the single-hotel walkthrough the counter-offer was ``a separately approved negotiation action,'' and with one supplier the approval could be a fixed rule such as a percentage below the opening offer. With three suppliers the approved price is a function of the other offers. A human can set it, which takes that step out of agent mediation. A policy can compute it from the airline's private state, in which case the released value is itself a channel. The interface fixtures of \cref{sec:research-agenda} contain the one-bit form of this: a work order whose room count is $20+s$, for a secret bit $s$, passes every schema and range check and delivers $s$ to the recipient. A counter-offer price is the many-bit form. \Cref{eq:method-release} guarantees that the outbound record is a deterministic function of the approved work order and public context; it says nothing when the work order's own values were chosen from private state. The release interface therefore holds for the initial request and for a fixed-rule counter, and stops where multi-supplier negotiation begins. Two known responses lie beyond it. The airline can accept a release that reveals a bounded amount, which needs a declassification policy stating what may be released and a measure of how much it reveals~\citep{sabelfeld2009declassification,smith2009qif}. Or it can replace bargaining with a mechanism such as a sealed-bid tender, in which the price a supplier is paid follows from rules announced in advance~\citep{vickrey1961}. Neither is part of the present design.

\paragraph{Who can reconstruct what:} \Cref{tab:example-multi} applies the reconstruction measure of \cref{sec:research-agenda} to each participant class, for the proposed method and for the comparator of signed point-to-point exchange with a directory and durable local histories.

\begin{table}[htbp]
\centering
\small
\caption{Reconstruction by participant class in the three-hotel case.}
\label{tab:example-multi}
\begin{tabularx}{\textwidth}{@{}P{3.6cm}LL@{}}
\toprule
Participant & Can retrieve and verify & Under the comparator \\
\midrule
Hotel A's agent, auditor or late-joining service & $d_A$ only: the request, offers, acceptance and its own terminal record & The same, from its bilateral log \\
Hotel B, Hotel C & Their own deal, as above & The same \\
Airline planner & $d_A$, $d_B$, $d_C$ and the private correlation & The same, from three bilateral logs and its own state \\
Airline finance, audit or logistics, joining later & $d_A$, $d_B$, $d_C$ as signed records, plus the signed correlation record, without the planner's process state & The same evidence if correlation is durably recorded and exposed to authorised services \\
\bottomrule
\end{tabularx}
\end{table}

For every hotel-side participant the permitted evidence is confined to one bilateral thread, as it is in a signed point-to-point log. This equivalence of authorised content does not imply equal retrieval latency, recovery time or integration cost. Inside the airline's boundary, a late-joining participant can retrieve all three deal threads and their private correlation without depending on the planner's surviving process state. Talea requires that correlation to be recorded explicitly; a point-to-point implementation can do the same. A fair test must retain and expose equivalent correlation evidence in both designs. Any remaining benefit must concern discovery, reconstruction effort or recovery performance, rather than evidence withheld from the comparator.

\paragraph{What the case shows:} Under the disclosure constraints of \cref{sec:method}, a multi-party case in which the counterparties compete decomposes into pairwise channels plus one hub that holds their union. The shared record converges on the point-to-point topology for every party except the hub, whatever substrate carries it. A possible benefit remains inside the hub's trust boundary, where participants are fluid and the planner's memory is not the record; the worked case does not demonstrate that benefit. The disclosure interface, meanwhile, covers fixed-value releases and stops at adaptive negotiation, which is the part of the exchange that reads private state by construction. The case does not examine counterparties who cooperate on one deal, such as a compliance or logistics service engaged by both sides, and who may agree to share its thread; the collapse is not claimed for them. \Cref{sec:findings} states these as the study's findings and restates the hypothesis within the boundary they draw.

\section{Findings, and what remains open}
\label{sec:research-agenda}
\label{sec:findings}

The study set out to test H1 and did not get that far. What it produced instead is an analysis of where a shared conversation record could support agent-mediated commerce and where confidentiality constrains it, drawn with an instrument that exists and a case worked through to its end. This section states the findings, the evidence behind each, and the narrower question that remains.

\subsection{What the study found}

\paragraph{Substrate fit:} A protocol built for federated social media provides mechanisms supporting the first five coordination requirements of \cref{sec:background-requirements} as stock infrastructure (\cref{tab:atproto-coordination}): verifiable identity, a signed attributable record, published schemas, binding to exact content, and discovery with fluid participation. Nothing had to be added to the protocol to obtain them; the custom Lexicons ride the same relays and indexes as social records. The sixth requirement, controlled access and release, is not provided at the record level. Relay crawl configuration restricts distribution by topology, and nothing restricts what an authorised publisher places in a record. Privacy on this substrate is a deployment property. Every confidentiality mechanism in \cref{sec:method}, the separate identities per level, the restricted deal channel, private correlation and the release interface, exists to supply what the protocol does not. The overlap has structural weight: both problems involve many parties, no shared operator and changing participation, and the one requirement the protocol lacks is the one a public-by-default network never needed. We wrote the requirements and chose the instrument ourselves, so the fit shows suitability and is not an independent test.

\paragraph{Collapse under confidentiality:} Where counterparties compete and may not see one another's terms, the multi-party record decomposes into bilateral threads plus one hub that holds their union (\cref{sec:example-multi}). Each hotel's authorised view contains exactly one thread, which is what a signed point-to-point log contains. The correlation that would make the record more than bilateral is the information that must not be published, so the shared record converges on the point-to-point topology for every party except the hub, whatever substrate carries it. This is not a weakness of the AT Protocol. A substrate with record-level access control would reach the same topology, because the constraint comes from the business, not from the protocol. The result is established for competing counterparties. Where organisations cooperate on one deal and agree to share its thread, as a buyer, a supplier and a compliance service might, that constraint is absent and the case examined here does not speak to it.

\paragraph{The hub residual:} A possible benefit remains inside the hub's trust boundary. There the record still holds three signed deal threads and the signed correlation record that links them, retrievable by a finance, audit or logistics participant that arrives after the fact, without the planner having been built to export its state (\cref{tab:example-multi}). A point-to-point design can also retain and expose this correlation. The comparison must therefore hold correlation evidence and authorised access equal, rather than attribute missing evidence to the transport. Where participants are fluid and the planner's memory is not the record, H1 could still hold. That is the setting the remaining hypothesis names.

\paragraph{The declassification boundary:} The release interface of \cref{eq:method-release} constructs an outbound message deterministically from an approved work order and public context, and the fixtures confirm that private context is not an input to it. It covers the initial request and a fixed-rule counter. It stops at the point where the work order's own values are chosen from private state, which is what a counter-offer against competing bids is. The one-bit fixture, a room count of $20+s$ that passes every schema check and carries $s$, and the many-bit counter-offer price are the same channel. This is where information-flow control meets planner-chosen values, and no field allowlist closes it. Bounded-leakage declassification policies and sealed-bid mechanisms are the known ways past it (\cref{sec:example-multi}); the design contains neither.

\paragraph{Composition:} The sandbox establishes that AT Protocol records and A2A execution compose. Published requests reach agents through triage and the bridge, responses return to the thread, and in a documented restart exercise an agent answered a follow-up from reconstructed context. Three properties of the substrate shaped the design and belong with the findings rather than with the caveats. There is no unsend: a published record has already flowed through the event stream, so release must be checked before publication, not repaired after. There is no global order: each repository orders its own commits, and timestamp order across repositories is presentation rather than causality, so dependency must be declared by reference. And C6 is a deployment property, so a deployment's confidentiality claims are claims about its topology and its gateways, never about the protocol.

\paragraph{Supporting checks:} The two-deal finite model explores atomic acceptance, seller confirmation or closure, delayed terminal delivery and resource constraints. Its three reference configurations contain 448, 448 and 576 reachable states with no violation of the stated invariants. Removing budget or inventory admission permits overcommitment; releasing funds at timeout permits an already confirmed obligation to lose its backing. The deterministic fixtures reject 13 altered mandate cases and construct identical permitted payloads across 18 work orders and 48 private-context variants per order (864 payload checks; 3,456 rejected candidate mutations). One fixture is a deliberate counterexample: a work order whose room count is $20+s$ passes a schema-only range check for either value of a secret bit $s$, so validating field domains does not bound what a planner-chosen value conveys. Both reproduce from \texttt{protocol\_model.py} and \texttt{boundary\_checks.py}. They exhibit failures when selected admission controls are removed and check the specified fixed-payload interface. The model abstracts from variable exposure, cryptographic verification and the full buyer-state machine; in particular, it does not implement the prohibition on acceptance after an observed closure. Its state counts therefore concern the supplied abstraction, not exhaustive verification of Algorithm~\ref{alg:method-admission}. They do not compare representations, and so do not measure a benefit of the shared record; nor do they test resistance to prompt injection, adaptive leakage or unauthorised recipient selection.

\subsection{What remains open}

H1 survives the map as a narrower claim than it was stated. For competing parties across a commercial boundary, the case establishes the same bilateral scope of authorised evidence under either design; performance differences remain unmeasured. The claim that remains is this.

\begin{quote}
\textbf{H1$'$ (bounded coordination benefit).} Inside one trust boundary, for participants who join after a decision or recover after an interruption, a shared attributable record improves the completeness and timeliness of decision reconstruction and recovery, compared with point-to-point exchange using equivalent authority, reservation and retention controls, including the same correlation evidence.
\end{quote}

A deal thread that cooperating organisations agree to share falls under H1 as first stated. The study did not examine it, and it stays open in that form. H1$'$ is still empirical and cannot be proved from the transaction rules. The decisive experiment would compare Talea with signed point-to-point coordination using the same agents, commercial profile, identity trust, gateways, ledgers, release policies and retained evidence. A directory, history retrieval and durable cross-deal correlation must be available to the same authorised participants in both designs; omitting them from the comparator would confound representation with missing functionality. Workloads should include stable bilateral negotiation, late participant arrival on the hub side, delegation across organisations, a deal shared among cooperating organisations, and the multi-supplier case of \cref{sec:example-multi}, measured per participant class so that counterparties and the hypothesised benefit inside the hub are evaluated separately.

For each workflow, an independent test oracle would enumerate the records required to establish its decision. Reconstruction coverage is the fraction of those required records that an authorised participant can retrieve and verify. Outcome recovery time runs from service resumption or the start of an outcome query until the participant establishes the correct durable terminal state. The experiment should also record unresolved outcomes, duplicate external effects, storage and network cost, and the work needed to integrate a new participant. Thresholds for acceptable cost and meaningful benefit should be fixed before the comparison.

Failures should be injected at the interfaces the design depends on: after local reservation but before publication, after publication but before acknowledgement, and after seller confirmation but before buyer observation. Duplicate delivery, closure arriving before acceptance, delayed indexing and unavailable historical records require separate cases. A correct recovery must preserve resource backing even when the evidence needed for progress is temporarily unavailable. Reporting successful agent answers alone would miss this distinction.

\subsection{Semantic and disclosure boundaries}

A reproducible commercial test first needs one complete domain profile. The lodging profile must define units, valid quantities, cancellation charges, rounding and a deterministic maximum-exposure calculation. Two independent validators should derive the same obligation from each conformance case and reject unsupported or unbounded liabilities. The illustrative AED~72,000 cap in the worked case avoids these ambiguities; it does not resolve them for general trade.

Disclosure evaluation must distinguish permitted business content from private state. Tests should vary private budgets and planning context while keeping approved work orders fixed, then attempt additional fields, value encoding, recipient changes and repeated invocations. Counterparty messages and agent descriptions should be treated as adversarial inputs. Passing a fixed-payload test supports that interface boundary only. Adaptive negotiation sits on the other side of the declassification boundary above, and needs policy that does not yet exist.

The interpretation of any positive result on H1$'$ must remain conditional. Trusted gateways and serialisable ledgers exclude dishonest resource accounting. Permanent loss of durable state is outside the current recovery model. Amendment accounting and competing replacement revisions require explicit semantics before they enter the supported lifecycle. These are boundaries of the study, and each determines which outcome a future experiment can legitimately establish.

\section{Discussion}
\label{sec:discussion}

We built Talea to find out whether a conversation can become a common source of evidence for business commitments across organisations. The answer has a shape. A conversation can become that evidence when its participants, dependencies and outcomes are published in a form that authorised parties can retain and verify, and the substrate supplies nearly everything needed to do so. But commercial confidentiality draws a line through the record, and on the far side of that line each party's view is bilateral. Point-to-point integration between competing counterparties is not a failure of imagination; it is what the confidentiality constraint produces from any substrate. The shared record may offer a benefit inside the boundary, where participants change and no single service's memory is the record.

The strongest competing explanation for any benefit inside the boundary is that the familiar enforcement mechanisms do the consequential work. Atomic reservations prevent overspending; unique terminal decisions prevent contradictory local outcomes; durable outboxes preserve decisions across publication failures. These properties can be implemented with other transports, and the finite-model results do not discriminate between the explanations. A demonstrated benefit must therefore concern evidence discovery, reconstruction or recovery after those mechanisms are held equal, which is how H1$'$ is stated.

The same records that support reconstruction create a disclosure obligation, and the substrate makes it unforgiving: a published record cannot be recalled, so release has to be right before publication. The release interface answers this for approved work orders with fixed values. It does not answer it for adaptive negotiation, and the three-hotel case shows why a stricter interface would not close the gap. The planner's choice of a value is the channel. Closing it means taking the choice away from the agent, through a human or a mechanism with announced rules, or accepting that every counter-offer releases some private state and bounding how much. Which of those a business prefers is a policy question the substrate cannot settle.

The value of the approach therefore depends on deployment in a way the study makes specific. Stable workflows within one organisation may not justify federation; a database and queue supply the context more simply. Competing counterparties obtain no broader authorised evidence from the shared record than from an equivalent signed bilateral log; operational differences remain unmeasured. Cooperating organisations that agree to share a deal thread were not examined. The setting that remains is a hub with fluid internal participation: planners, finance, audit and logistics services that come and go, and that must reconstruct decisions made by services no longer running. An optional runtime such as Agent Substrate changes execution and isolation but not this analysis, and its effect should be evaluated separately from the record representation.

\section{Conclusion}
\label{sec:conclusion}

We used the AT Protocol as an instrument to find out where a shared, attributable conversation record helps agent-mediated business, and where it stops. It provides mechanisms supporting the first five coordination requirements; the missing one, controlled release, shapes everything built above it. Under commercial confidentiality between competitors the record collapses to pairwise channels for every party except the hub, and a possible benefit remains inside the hub's trust boundary. A deterministic release interface covers fixed-value disclosure and ends where adaptive negotiation begins. The sandbox shows the composition works; the model checks exhibit failures when selected transaction controls are removed.

The hypothesis that remains is narrower than the one we began with and can be tested with the experiment stated in \cref{sec:findings}. Whether a conversation should become part of the infrastructure for agent-mediated business is, on this evidence and for counterparties that compete, a question about the inside of an organisation more than the space between them.

\section*{Code and evidence availability}
The finite model (\texttt{protocol\_model.py}), deterministic interface fixtures (\texttt{boundary\_checks.py}), recorded JSON results and reproduction instructions accompany this submission as ancillary files. Both scripts use only the Python standard library; running them regenerates the reported results. These files reproduce the abstract model and interface checks, not the coordination sandbox or a deployed commercial system. The project repository is \url{https://github.com/carloseberhardt/talea/}; access to that repository is not required to reproduce the archived checks. The sandbox restart exercise is an author-reported observation documented in the project's \texttt{IMPLEMENTATION.md}; the complete sandbox is not included in the ancillary package. No business transaction data or participant data were used.

\section*{Use of AI assistance}
The authors used AI coding and writing assistants in building the sandbox and in drafting, editing and reviewing this manuscript. The authors reviewed all content and are responsible for it.

\bibliographystyle{unsrtnat}
\bibliography{references}

\clearpage
\appendix
\section{Admission procedure and safety properties}
\label{app:safety}

This appendix gives the gateway procedure and the three safety properties summarised in \cref{sec:method}. The symbols are those of \cref{eq:method-budget,eq:method-stock}.

\begin{algorithm}[H]
\caption{Gateway admission, publication and buyer reconciliation}
\label{alg:method-admission}
\KwIn{authenticated caller, action, offer $o$, deal revision $(d,v)$}
\KwState{authoritative policy and resource ledgers; durable action and terminal slots; buyer state; publication outbox}
\textbf{Admission at the acting organisation}\;
Resolve the caller's permitted signer; reject identity mismatch or malformed references\;
Begin a serialisable local transaction\;
\If{this action already has a durable outcome}{return the stored outcome without another resource update\;}
Verify the profile, exact terms, action authority and permitted release\;
\uIf{action is acceptance}{
Reject if buyer state is closed or committed\;
Check Eq.~\ref{eq:method-budget}; reject if admission fails\;
Reserve $E(o)$; set buyer state to reserved; store acceptance and authority receipt\;
}
\uElseIf{action is confirmation}{
Verify acceptance, exact offer binding and buyer authority evidence\;
\If{a seller terminal decision already exists}{return that decision\;}
\uIf{deadline permits confirmation and Eq.~\ref{eq:method-stock} holds}{
Allocate inventory; store confirmation in the terminal slot\;
}\Else{store closure in the terminal slot\;}
}
\uElseIf{action is closure}{
\If{a seller terminal decision already exists}{return that decision\;}
Store closure in the terminal slot; allocate no inventory\;
}\Else{abort and reject\;}
Store exact publication bytes and a stable record key in the outbox; commit\;
\textbf{Publication worker:} publish stored bytes; on retry, verify existing content at the same key; escalate conflicts\;
\BlankLine
\textbf{On receipt of a seller terminal record at the buyer}\;
Verify signer, deal, revision and required reference bindings\;
Begin a serialisable local transaction\;
\If{this terminal outcome is already applied}{return without accounting changes\;}
\uIf{record confirms and buyer state is reserved}{
Atomically move $E(o)$ from reserved to committed exposure; set state to committed\;
}
\uElseIf{record closes and buyer state is unseen or reserved}{
Release any reservation; set state to closed, preventing later acceptance\;
}\Else{retain current accounting; quarantine the inconsistent outcome\;}
For a consistent outcome, persist its reconciliation marker with the state change; commit\;
\textbf{Timeout:} retain reservation and query the terminal outcome; silence releases no funds\;
\end{algorithm}

The following propositions concern initial deal revisions with fixed, non-negative exposure. They assume authenticated records, trusted gateways, durable serialisable ledgers, and enforcement by every writer of the controlled resources. Budget $B$ is fixed during the trace. Each revision has one durable buyer state, either unseen, reserved, committed or closed. Reconciliation atomically updates this state and its accounting entries. A verified closure received before acceptance records a closed state and prevents subsequent acceptance of that revision; repeated messages do not change an already reconciled state. Amendments, settlement, loss of durable state and third-party cancellation are outside these proofs.

\begin{proposition}[Budget preservation]
\label{prop:budget-safety}
Let $U=C+\sum_{r\in R}E(r)$ denote accounted exposure. If $U\leq B$ initially, and every acceptance is admitted atomically only when $U+E(o)\leq B$, then $U\leq B$ after every transition of the initial-commitment protocol.
\end{proposition}
\begin{proof}
Proceed by induction over the serialised buyer transitions. An admitted acceptance adds $E(o)$ to the reserved exposure only after checking the bound. Confirmation reconciliation removes that amount from reservations and adds the same amount to committed exposure, leaving $U$ unchanged. Closure reconciliation removes a non-negative reservation. A closure recorded before acceptance changes no exposure and prevents later acceptance. Publication, timeout and repeated delivery change neither accounting entry. Each transition therefore preserves the bound, completing the induction.
\end{proof}

\begin{proposition}[Exclusive terminal decision]
\label{prop:terminal-exclusivity}
If the seller admits confirmation or closure only by an atomic transition of the same durable decision slot from undecided to a terminal state, and terminal states are irreversible, then one revision cannot have both an admitted confirmation and an admitted closure.
\end{proposition}
\begin{proof}
Serialisability orders competing attempts on the slot. The first successful transition replaces undecided with one terminal state. Every subsequent attempt observes a terminal state and returns the existing decision without replacing it. Thus the other terminal transition cannot succeed. Publication retries reproduce the stored decision and introduce no additional transition.
\end{proof}

\begin{proposition}[Backing during delayed communication]
\label{prop:reservation-backing}
Suppose the buyer durably reserves $E(o)$ before publishing acceptance, the seller confirms only that exact accepted revision, and the buyer releases the reservation only on verified seller closure. During the initial-commitment lifecycle, every seller-confirmed revision retains its exposure in either the buyer's reservation or its committed accounting entry, despite message delay or duplication.
\end{proposition}
\begin{proof}
Confirmation requires an acceptance whose publication follows durable reservation. By Proposition~\ref{prop:terminal-exclusivity}, a confirmed revision cannot also have a valid seller closure. The buyer therefore cannot take the release transition for that revision. Before confirmation is observed, the reservation remains; after atomic reconciliation, its value is transferred to committed exposure. Timeout and duplicate messages do not alter either conclusion. Backing is preserved, although an unavailable terminal record can prevent progress indefinitely.
\end{proof}

\end{document}